\documentclass{article}
\usepackage{ijcai20}

\usepackage{times}

\usepackage{soul}
\usepackage{url}
\usepackage[utf8]{inputenc}
\usepackage[small]{caption}
\usepackage{graphicx}
\usepackage{amsmath}
\usepackage{booktabs}
\usepackage{xcolor}
\usepackage{float}

\usepackage{soul} 

\title{Considerations, Good Practices, Risks and Pitfalls\\ in Developing AI Solutions Against COVID-19}

\author{
Alexandra Luccioni$^{1,2}$\and
Joseph Bullock$^{3,4}$\and
Katherine Hoffmann Pham$^{3,5}$\and \\ 
Cynthia Sin Nga Lam$^6$\And
Miguel Luengo-Oroz$^3$
\affiliations
$^1$~Universit\'{e} de Montr\'{e}al  
$^2$~Mila Qu\'{e}bec Artificial Intelligence Institute\\
$^3$~United Nations Global Pulse 
$^4$~Institute for Data Science, Durham University\\ 
$^5$~NYU Stern School of Business 
$^6$~Global Coordination Mechanism on NCDs, WHO
\emails
sasha.luccioni@mila.quebec, \{joseph,katherine,miguel\}@unglobalpulse.org,
lams@who.int}

\begin{document}

\maketitle

\begin{abstract}
The COVID-19 pandemic has been a major challenge to humanity, with 12.7 million confirmed cases as of July 13th, 2020~\cite{who2020}. In previous work, we described how 
Artificial Intelligence can be used to tackle the pandemic with applications at the molecular, clinical, and societal scales~\cite{bullock2020mapping}. In the present follow-up article, we review these three research directions, and assess the level of maturity and feasibility of the approaches used, as well as their potential for operationalization. 
We also summarize some commonly encountered risks and practical pitfalls, as well as guidelines and best practices for formulating and deploying AI applications at different scales.   
\end{abstract}

\begin{figure*}[h!]
\centering
\includegraphics[width=.8\textwidth]{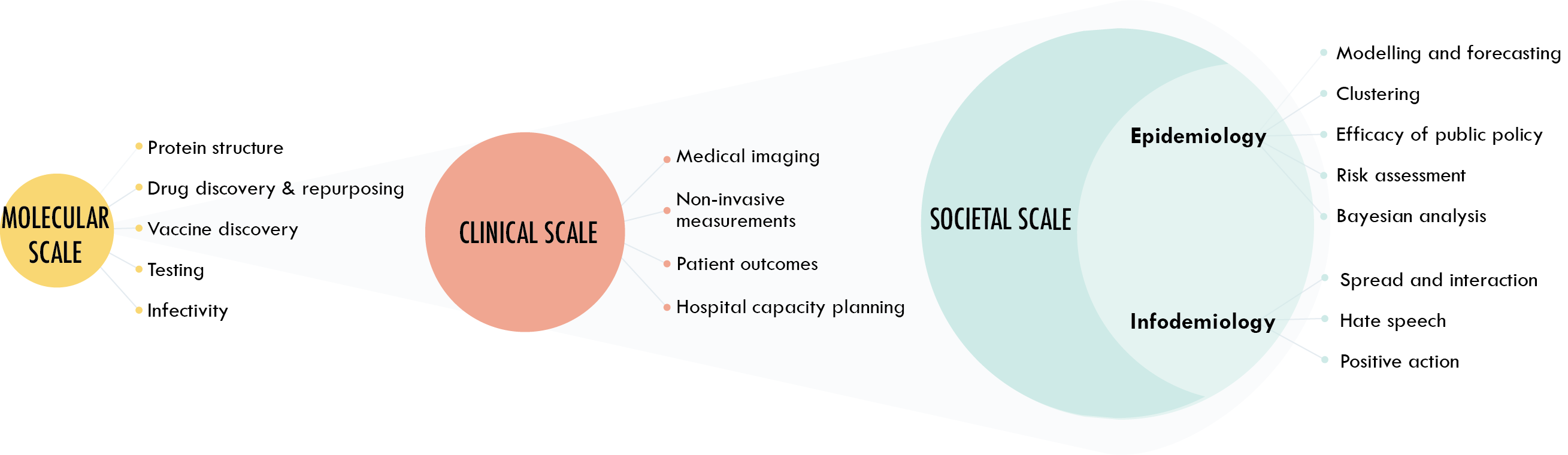}
\caption{AI applications for the COVID-19 response organized at three levels: the molecular scale, the clinical scale, and the societal scale. }\label{fig:scales}
\end{figure*}

\section{Introduction}

In recent months, an already remarkable body of research on 
potential applications of Artificial Intelligence (AI) and Machine Learning (ML) to address the health and societal challenges created by the COVID-19 pandemic has been produced. In a recent survey of the literature~\cite{bullock2020mapping}, we identified over 300 relevant journal articles and preprints with applications ranging from tracking the spread of misinformation to molecular modeling of the SARS-CoV-2 virus. These applications are emerging in a wide array of settings, increasingly driven by the priorities of responders and multidisciplinary collaborations.

In this article, we present a more in-depth analysis of AI applications at the molecular, clinical, and societal scales. We build on our previous work by sharing some high-level observations on the broader landscape of AI for COVID-19 and providing a critical take on best practices and application strategies. While the speed and scale of the AI community's research response to the pandemic is unprecedented, we feel nonetheless that there are a number of common, structural challenges faced by many of these projects. We extract shared design considerations and promising directions, as well as recurring problems, risks, and areas of improvement that could help researchers and funding bodies when developing further applications and research agendas. 

\section{Molecular Scale}
\paragraph{Promising Directions}

AI has been applied in a wide spectrum of molecular research -- ranging from better understanding the structure of the SARS-CoV-2 virus to assisting drug development and improving molecular diagnosis.

Some of the earliest responses to the pandemic were produced by AI researchers leveraging existing protein structure prediction algorithms \cite{jumper_computational_2020} and drug discovery pipelines \cite{zhavoronkov_potential_2020}, or exploring pre-built knowledge graphs \cite{richardson_baricitinib_2020}. Having the infrastructure for this research already in place facilitated a rapid response to the new challenge of COVID-19. Open science, based on previously published peer-reviewed papers detailing the methodologies involved, has helped to accelerate the evaluation of these approaches.

A number of projects studied drug re-purposing, in the hope of discovering a therapy for which the engineering of novel compounds is not needed, clinical trials have been authorized, or the compound's use has already been approved (e.g., 
\cite{nguyen_potentially_2020}).
 Others used ML approaches to reduce the computational burden of docking simulations by narrowing down the set of candidate compounds which needed to be docked (e.g., \cite{ton_rapid_2020}). 
Another notable trend in molecular applications involves identifying desirable or undesirable properties of a candidate compound -- e.g., novelty, drug likeness, or toxicity -- and then training drug discovery models to suggest compounds that meet these desiderata (e.g., \cite{tang_ai-aided_2020}). 
These applications allowed creative exercises from the modeling perspective, in which researchers experimented with different representations of the same data to design different AI pipelines.

Many papers relied on a common set of open datasets such as PDBbind, ChemBL, and DrugBank, which makes their methodologies more accessible and replicable for other researchers. Several even posted models or results on Github or other web pages, providing open access to all (e.g., \cite{lopez-rincon_accurate_2020}).
\paragraph{Risks and Pitfalls}

Many ML applications in molecular science remain at the research level. It is understandable that molecular research is part of a multidisciplinary and multi-step process; nonetheless, few of the efforts we observed had advanced through the drug development pipeline to more formal evaluation. 
One challenge is that synthesizing and testing compounds is costly and time-consuming, so researchers who are not embedded within a larger infrastructure are often unable to execute these later steps required to translate research to practice. 

Another barrier in this research-to-practice process could be limitations in data and model sharing. To our knowledge, two candidate vaccines that reported the use of ML in their development have been approved for clinical evaluation \cite{who2020vaccines}. However, both vaccines came from corporations which published limited information on their approach and the extent to which ML was utilized in vaccine development. More generally, we found that some of the research produced by private-sector entities relied on datasets such as knowledge graphs which are not publicly accessible, describing proprietary models only in vague or high-level terms.
Hence, we would like to reiterate the importance of open science in ensuring accessible vaccines and treatments for vulnerable communities.

\section{Clinical Scale}

\paragraph*{Promising Directions}
From a clinical perspective, AI and ML have already been used to assist and improve patient-level assessment of COVID-19. One major application for ML in this field has been the analysis of medical imagery such as CT and X-Ray scans with the help of common neural network architectures, either to provide a supplementary data point to corroborate COVID-19 diagnosis~\cite{li2020artificial}, or to assess the severity and progression of the disease~\cite{tang2020severity}. Some of these approaches are now fully operational and have received institutional approval for deployment in hospitals as triage tools~\cite{alaa2018autoprognosis} or as human-in-the-loop systems for radiologists~\cite{shan2020lung}.

Approaches using complementary data sources have also been proposed, ranging from wearable devices and mobile phones for detecting symptoms~\cite{radin2020harnessing}, to electronic health records to improve diagnosis and outcome prediction. We find one of the most promising applications of AI at a clinical level to be the prediction of patient outcomes and the proposal of triage approaches based on features extracted from medical data. Such approaches are transparent and practicable, pinpointing key measurable features which enable hospitals to plan the use of resources such as ventilators and ICU beds.

Furthermore, there are promising hybrid studies that leverage both medical imagery and clinical features to predict patient-level characteristics such as the severity of COVID-19~\cite{shi2020deep}. These approaches leverage different complementary sources of data in order to make more precise and more generalizable predictions of patients' prognosis. In fact, the clinical features identified by studies using solely clinical data were also corroborated by hybrid studies; e.g. both sets of studies found that high levels of substances such as lactic dehydrogenase (LDH) and high-sensitivity C-reactive protein (CRP) were correlated with  mortality risk and longer hospitalization.  

\paragraph{Risks and Pitfalls}
While many studies are being carried out in situ on COVID-19 patients, there is still much that we do not know about the virus itself and the factors that can put patients at risk for hospitalization, developing acute respiratory distress syndrome (ARDS), and eventual death from respiratory failure. For instance, the extent to which medical imagery alone can be used for the diagnosis of COVID-19 is still debated by the medical community ~\cite{weinstock2020}. Above and beyond the feasibility of diagnosis, many of the medical imaging papers we reviewed had methodological issues, relying on small and poorly-balanced datasets that mix data from several populations, coupled with flawed evaluation procedures~\cite{wynants2020prediction}. Most also presented no plan for inclusion in clinical workflows and no attempt to provide a transparent explanation for the 
diagnosis, which is especially important in patient-level applications of AI~\cite{wiens2019no}. Finally, while approaches that leverage ML to analyze non-invasive measurements are potentially promising given the ubiquity and accessibility of sensor technologies, we found that these approaches are not sufficiently mature to evaluate their performance. We would advocate more extensive testing and clinical investigations to validate their performance in deployment. 

\section{Societal Scale}

\paragraph{Promising Directions}

From a societal perspective, AI has been applied to the field of epidemiological modelling, as well as to understanding and combating the ``infodemic'' spread of misinformation~\cite{who2020infordmic}. At the epidemiological level, many studies have sought to produce forecasting models for national and regional level statistics. While a vast body of literature on modeling already exists, AI based models could 
augment classical models in situations where analytic transmission equations are not well known, such as when modelling the effects of public policy measures such as social distancing and self-quarantine~\cite{dandekar2020neural}. In addition, AI methods can also be used to incorporate 
new data sources, such as social media and search information~\cite{lampos2020tracking}. Other works use AI to identify similarities and differences in the evolution of the pandemic between regions. These approaches have leveraged both supervised and unsupervised techniques, and may help inform policy makers at a high level and highlight areas for more detailed exploration. 

AI has also been applied to investigate the scale and spread of the infodemic in order to address the propagation of misinformation and disinformation including the emergence of hate speech. Given the vast amount of information now being disseminated and shared, there is a need for tools to help identify and promote reliable information sources, and understand the spread of misinformation. Promising work has analyzed patterns in the transmission of such information and developed infodemic risk scoring algorithms (e.g., \cite{gallotti2020assessing}). Moreover, there has been an increasing focus on assessing the emergence of hate speech, particularly using network analysis techniques, which could help inform preventative efforts or contribute to the development of 
continuous monitoring platforms~\cite{velasquez2020hate}.

\paragraph{Risks and Pitfalls}
Policy decisions must be based on justifiable models which stand up to public scrutiny. Since much of the data collected for COVID-19 epidemiological modeling tasks is extremely limited, the choice of models and datasets can have significant effects on overall performance and models may lack generalizability. A significant limitation of many articles in this category is the heterogeneous data collection in different countries due to multiple factors including variations in testing, case tracking, and reporting quality and standards. Moreover, applying models trained in one context to another raises concerns surrounding the model's ability to capture aspects such as different cultural norms which may impact 
the spread and effect of the virus. In such examples, a transferred model will have to be tailored for local contexts given that there may be different demographic characteristics and behaviors. Indeed, developing proper model transferability procedures and guidelines is especially important for data poor regions. While synthetic data approaches have been proposed, these should be applied 
with caution.

In order to understand and tackle the infodemic,
it is important to capture and analyze information from a diverse range of sources. While much of the work on this topic uses data from online sources such as social media and Google searches, information propagated through alternative channels such as radio is important for capturing wider trends. Moreover, many of the existing approaches rely on language modelling techniques developed for English or other widely spoken languages, but relying on such models might leave many populations behind, including some of those most vulnerable. Finally, we note that while numerous methods have been proposed for identifying hate speech, further research is needed to identify the targeted groups and to learn how to use the insights gained from these techniques in an effective way (e.g. see Section II \cite{unhatespeech}).


\section{Discussion}

We believe that there are several considerations to keep in mind when applying AI to a global problem such as the COVID-19 pandemic. These include:
\begin{enumerate}
\item \emph{Application relevance and context}: Does the application make sense from both an application and a methodological perspective? Were domain experts consulted to properly assess the needs and define the problem to be solved? Does the solution serve its target audience? 
\item \emph{Data availability and quality}: Has the quantity of data being used to train and evaluate the model been assessed and deemed to be of a reasonable size and diversity to justify the claims made? Has bias in the data been considered and documented? Were privacy measures taken?
\item \emph{AI methodology and complexity}: Is the approach proposed justified? If an AI based model, has it been benchmarked against more traditional approaches? Has the approach been validated by other researchers and, if possible, peer-reviewed?
\item \emph{Transparency and explainability}: Have efforts been made to render the results and the approach understandable by humans? Is it possible to identify the key features being used by the algorithm?
\item \emph{Dissemination of knowledge}: Are the data, code and models being shared in any form? Are there reporting guidelines and standards that should be followed? 
\item \emph{Operationalization and performance}: Can the approach be integrated into decision-making workflows? What is the implementation plan? Are there regulatory frameworks that have to be taken into account? What are the potential risks? Will the models incorporate user feedback, and if so, how?
\end{enumerate}

The COVID-19 pandemic is a global emergency that has overstretched health care networks and posed significant health, economic and social challenges to humanity. AI can play an important role in alleviating this pressure, but we would like to reiterate that in order for any technological solution to make an impact, it must be deployed contextually and appropriately. We advocate for aspiring research initiatives to be carried out in partnership with stakeholders who have the necessary domain knowledge. 
It is also important to investigate how proven solutions can be adapted to local contexts to address unmet needs, particularly in areas of the world with fewer resources. This requires developing appropriate model and data sharing solutions, and specific measures to address data scarcity (see e.g.~\cite{wright2020vulnerable} for more details).

Finally, given the rapidly changing nature of human understanding regarding the pandemic, and therefore the inability to fully validate many approaches, we suggest that models should not be designed to process data in an end-to-end fashion at this stage, but rather to augment human decision making. With careful attention to the implementation context and operational needs, we believe that AI solutions can be a valuable asset in the fight against the pandemic.

\section*{Acknowledgements}

United Nations Global Pulse is supported by the Governments of Netherlands, Sweden and Germany and the William and Flora Hewlett Foundation. JB also is supported by the UK Science and Technology Facilities Council (STFC) grant number ST/P006744/1. AL is supported by funding from IVADO and Mila.

\bibliographystyle{plos2015}
\bibliography{main_IJCAI}

\begin{thebibliography}{10}

\bibitem{who2020}
WHO. Coronavirus disease ({COVID}-19) outbreak situation; 2020.
\newblock
  \url{https://www.who.int/emergencies/diseases/novel-coronavirus-2019/situation-reports}.

\bibitem{bullock2020mapping}
Bullock J, Luccioni A, Pham KH, Lam CSN, Luengo-Oroz M, et~al.
\newblock Mapping the landscape of artificial intelligence applications against
  {COVID}-19.
\newblock arXiv preprint arXiv:200311336. 2020;.

\bibitem{jumper_computational_2020}
Jumper J, Tunyasuvunakool K, Kohli P, Hassabis D, {AlphaFold Team}.
  Computational predictions of protein structures associated with {COVID}-19;
  2020.

\bibitem{zhavoronkov_potential_2020}
Zhavoronkov A, Aladinskiy V, Zhebrak A, Zagribelnyy B, Terentiev V, Bezrukov
  DS, et~al.
\newblock Potential {{COVID}}-2019 {{3C}}-like {{Protease Inhibitors Designed
  Using Generative Deep Learning Approaches}}.
\newblock Insilico Medicine Hong Kong Ltd A. 2020;307:E1.

\bibitem{richardson_baricitinib_2020}
Richardson P, Griffin I, Tucker C, Smith D, Oechsle O, Phelan A, et~al.
\newblock Baricitinib as Potential Treatment for 2019-{{nCoV}} Acute
  Respiratory Disease.
\newblock The Lancet. 2020;395(10223):e30--e31.

\bibitem{nguyen_potentially_2020}
Nguyen DD, Gao K, Chen J, Wang R, Wei G.
\newblock Potentially highly potent drugs for 2019-{nCoV}.
\newblock bioRxiv preprint bioRxiv:20200205936013v1. 2020;.

\bibitem{ton_rapid_2020}
Ton AT, Gentile F, Hsing M, Ban F, Cherkasov A.
\newblock Rapid {{Identification}} of {{Potential Inhibitors}} of
  {{SARS}}-{{CoV}}-2 {{Main Protease}} by {{Deep Docking}} of 1.3 {{Billion
  Compounds}}.
\newblock Molecular Informatics. 2020;39:1--8.

\bibitem{tang_ai-aided_2020}
Tang B, He F, Liu D, Fang M, Wu Z, Xu D.
\newblock {{AI}}-Aided Design of Novel Targeted Covalent Inhibitors against
  {{SARS}}-{{CoV}}-2.
\newblock bioRxiv preprint bioRxiv:20200303972133. 2020;.

\bibitem{lopez-rincon_accurate_2020}
{Lopez-Rincon} A, Tonda A, {Mendoza-Maldonado} L, Claassen E, Garssen J,
  Kraneveld AD.
\newblock Accurate Identification of {SARS-CoV-2} from Viral Genome Sequences
  Using Deep Learning.
\newblock bioRxiv preprint bioRxiv:20200313990242v1. 2020;.

\bibitem{who2020vaccines}
WHO. Draft landscape of {COVID}-19 candidate vaccines; 2020.
\newblock
  \url{https://www.who.int/who-documents-detail/draft-landscape-of-covid-19-candidate-vaccines}.

\bibitem{li2020artificial}
Li L, Qin L, Xu Z, Yin Y, Wang X, Kong B, et~al.
\newblock Artificial intelligence distinguishes {COVID}-19 from community
  acquired pneumonia on chest {CT}.
\newblock Radiology. 2020;.

\bibitem{tang2020severity}
Tang Z, Zhao W, Xie X, Zhong Z, Shi F, Liu J, et~al.
\newblock Severity Assessment of Coronavirus Disease 2019 ({COVID}-19) Using
  Quantitative Features from Chest {CT} Images.
\newblock arXiv preprint arXiv:200311988. 2020;.

\bibitem{alaa2018autoprognosis}
Alaa AM, van~der Schaar M.
\newblock Autoprognosis: Automated clinical prognostic modeling via bayesian
  optimization with structured kernel learning.
\newblock arXiv preprint arXiv:180207207. 2018;.

\bibitem{shan2020lung}
Shan F, Gao Y, Wang J, Shi W, Shi N, Han M, et~al.
\newblock Lung Infection Quantification of {COVID}-19 in {CT} Images with Deep
  Learning.
\newblock arXiv preprint arXiv:200304655. 2020;.

\bibitem{radin2020harnessing}
Radin JM, Wineinger NE, Topol EJ, Steinhubl SR.
\newblock Harnessing wearable device data to improve state-level real-time
  surveillance of influenza-like illness in the USA: a population-based study.
\newblock The Lancet Digital Health. 2020;.

\bibitem{shi2020deep}
Shi W, Peng X, Liu T, Cheng Z, Lu H, Yang S, et~al.
\newblock Deep Learning-Based Quantitative Computed Tomography Model in
  Predicting the Severity of {COVID}-19: A Retrospective Study in 196 Patients.
\newblock arXiv preprint. 2020;.

\bibitem{weinstock2020}
Weinstock M, Echenique A, Russell Jea.
\newblock Chest {X}-ray findings in 636 ambulatory patients with {COVID}-19
  presenting to an urgent care center: {A} normal chest {X}-ray is no
  guarantee.
\newblock The Journal of Urgent Care Medicine. 2020; p. 13--18.

\bibitem{wynants2020prediction}
Wynants L, Van~Calster B, Bonten MM, Collins GS, Debray TP, De~Vos M, et~al.
\newblock Prediction models for diagnosis and prognosis of {COVID}-19
  infection: {S}ystematic review and critical appraisal.
\newblock BMJ. 2020;369.

\bibitem{wiens2019no}
Wiens J, Saria S, Sendak M, Ghassemi M, Liu VX, Doshi-Velez F, et~al.
\newblock Do no harm: a roadmap for responsible machine learning for health
  care.
\newblock Nature medicine. 2019;25(9):1337--1340.

\bibitem{who2020infordmic}
WHO. Infodemic management - Infodemiology; 2020.
\newblock
  \url{https://www.who.int/teams/risk-communication/infodemic-management}.

\bibitem{dandekar2020neural}
Dandekar R, Barbastathis G.
\newblock Neural Network aided quarantine control model estimation of COVID
  spread in {W}uhan, {C}hina.
\newblock arXiv preprint arXiv:200309403. 2020;.

\bibitem{lampos2020tracking}
Lampos V, Moura S, Yom-Tov E, Edelstein M, Majumder M, Hamada Y, et~al.
\newblock Tracking {COVID}-19 using online search.
\newblock arXiv preprint arXiv:200308086. 2020;.

\bibitem{gallotti2020assessing}
Gallotti R, Valle F, Castaldo N, Sacco P, Domenico MD.
\newblock Assessing the risks of ``infodemics” in response to {COVID}-19
  epidemics.
\newblock arXiv preprint arXiv:200403997. 2020;.

\bibitem{velasquez2020hate}
Velásquez N, Leahy R, Restrepo NJ, Lupu Y, Sear R, Gabriel N, et~al.
\newblock Hate multiverse spreads malicious {COVID}-19 content online beyond
  individual platform control.
\newblock arXiv preprint arXiv:200400673. 2020;.

\bibitem{unhatespeech}
{United Nations}. United {N}ations Guidance Note on Addressing and Countering
  {COVID}-19 related Hate Speech; 2020.
\newblock
  \url{https://www.un.org/en/genocideprevention/documents/Guidance%20on%20COVID-19%20related%20Hate%20Speech.pdf}.

\bibitem{wright2020vulnerable}
Wright J, Verity A.
\newblock Artificial Intelligence Principles For Vulnerable Populations in
  Humanitarian Contexts.
\newblock Digital Humanitarian Network. 2020;.

\end{thebibliography}

\end{document}